\documentclass[sigconf]{acmart}

\usepackage{booktabs} 
\usepackage{tikz}
\usepackage{tikz-uml}
\usepackage{amsmath}
\usepackage{listings}

\setcopyright{rightsretained}

\acmDOI{10.475/123_4}

\acmISBN{123-4567-24-567/08/06}

\acmConference[SC17]{SuperComputing}{November 2017}{Denver, Colorado, USA} 
\acmYear{2017}
\copyrightyear{2017}

\acmPrice{15.00}
\definecolor{darkgreen}{rgb}{0.33, 0.42, 0.18}
\definecolor{mauve}{rgb}{0.88, 0.69, 1.0}
\begin{document}
\title{High-level python abstractions for optimal checkpointing in inversion problems}

\author{Navjot Kukreja}
\authornote{Corresponding Author. Email: nkukreja@imperial.ac.uk}
\affiliation{%
  \institution{Imperial College London}
  \city{London}
  \country{UK}}
\author{Jan H\"uckelheim}
\affiliation{%
  \institution{Imperial College London}
  \city{London}
  \country{UK}}
\author{Michael Lange}
\affiliation{%
  \institution{Imperial College London}
  \city{London}
  \country{UK}}
\author{Mathias Louboutin}
	\affiliation{%
	\institution{The University of British Columbia}
	\city{Vancouver, BC}
	\country{Canada}}
\author{Andrea Walther}
\affiliation{%
  \institution{Universit\"at Paderborn}
  \city{Paderborn}
  \country{Germany}}
\author{Simon W. Funke}
\affiliation{%
  \institution{Simula Research Laboratory}
  \city{Lysaker}
  \country{Norway}}
\author{Gerard Gorman}
\affiliation{%
  \institution{Imperial College London}
  \city{London}
  \country{UK}}
\renewcommand\shortauthors{Kukreja, N. et al}

\newcommand{\simon}[1]{{\color{red}{Simon: \textbf{#1}}}}
\newcommand{\revolve}{\emph{Revolve} }
\newcommand{\crevolve}{\emph{cRevolve} }
\newcommand{\pyrevolve}{\emph{pyRevolve} }

\begin{abstract} 
Inversion and PDE-constrained optimization problems often rely on solving the
adjoint problem to calculate the gradient of the objective function. This
requires storing large amounts of intermediate data, setting a limit
to the largest problem that might be solved with a given amount of
memory available. Checkpointing is an approach that can reduce the
amount of memory required by redoing parts of the computation instead
of storing intermediate results. The Revolve checkpointing algorithm
offers an optimal schedule that trades computational cost for smaller
memory footprints. Integrating Revolve into a modern python HPC code
and combining it with code generation is not straightforward. We
present an API that makes checkpointing accessible from a DSL-based
code generation environment along with some initial performance figures
with a focus on seismic applications. 
\end{abstract}

%
%

\ccsdesc[500]{Software and its engineering~Software design engineering}
%
%

\keywords{HPC, Code generation, API, Checkpointing, Adjoint, Inverse Problems}

\maketitle

\section{Introduction}
Seismic inversion is a computationally intensive technique that uses
data from seismic wave propagation experiments to estimate physical
parameters of the earth's subsurface. A seismic inversion problem
based on a wave equation can be viewed as an optimization problem and numerically
solved using a gradient-based optimization 
\cite{virieux2009overview}. Since the gradient is usually calculated
using the adjoint-state method, the method requires that the forward
and adjoint field are known for each time step of the simulation
\cite{plessix2006review}. We discuss this in section
\ref{sec:inversion_devito}.

Previous work on similar inverse problems led to the \revolve
algorithm \cite{griewank2000} and the associated C++ tool which
provides an optimal schedule at which to store checkpoints,
i.e. states from which the forward simulation can be restored. A study
of optimal checkpointing for seismic inversion was done in
\citep{SymesRTM} but this was not accompanied by a high-level
abstraction that made integration of other software easier with
Revolve. The algorithm is further discussed in section \ref{sec:revolve}.

The \revolve tool and algorithm, however, only provide the schedule to be
used for checkpointing. Although this eases some of the complexity of
the application code using the algorithm, the glue code required to
manage the forward and adjoint runs is still quite complex. This acts
as a deterrent to the more widespread use of the algorithm in the
community. 

In this paper, we describe how the \revolve algorithm can be combined with code
generation to make checkpointing much more accessible. The software that
can enable this is described in section \ref{sec:api}. Although we use
particular examples from seismic imaging, the abstraction and software
proposed here are quite general in nature and can be used in
any problem that requires checkpointing in combination with a variety
of computational methods.  

In section \ref{sec:experiment} we provide some initial performance
figures on which we judged the correctness and performance of the
implementation. 

\section{Seismic imaging and Devito}
\label{sec:inversion_devito}
\begin{figure}
\includegraphics[width=0.8\linewidth]{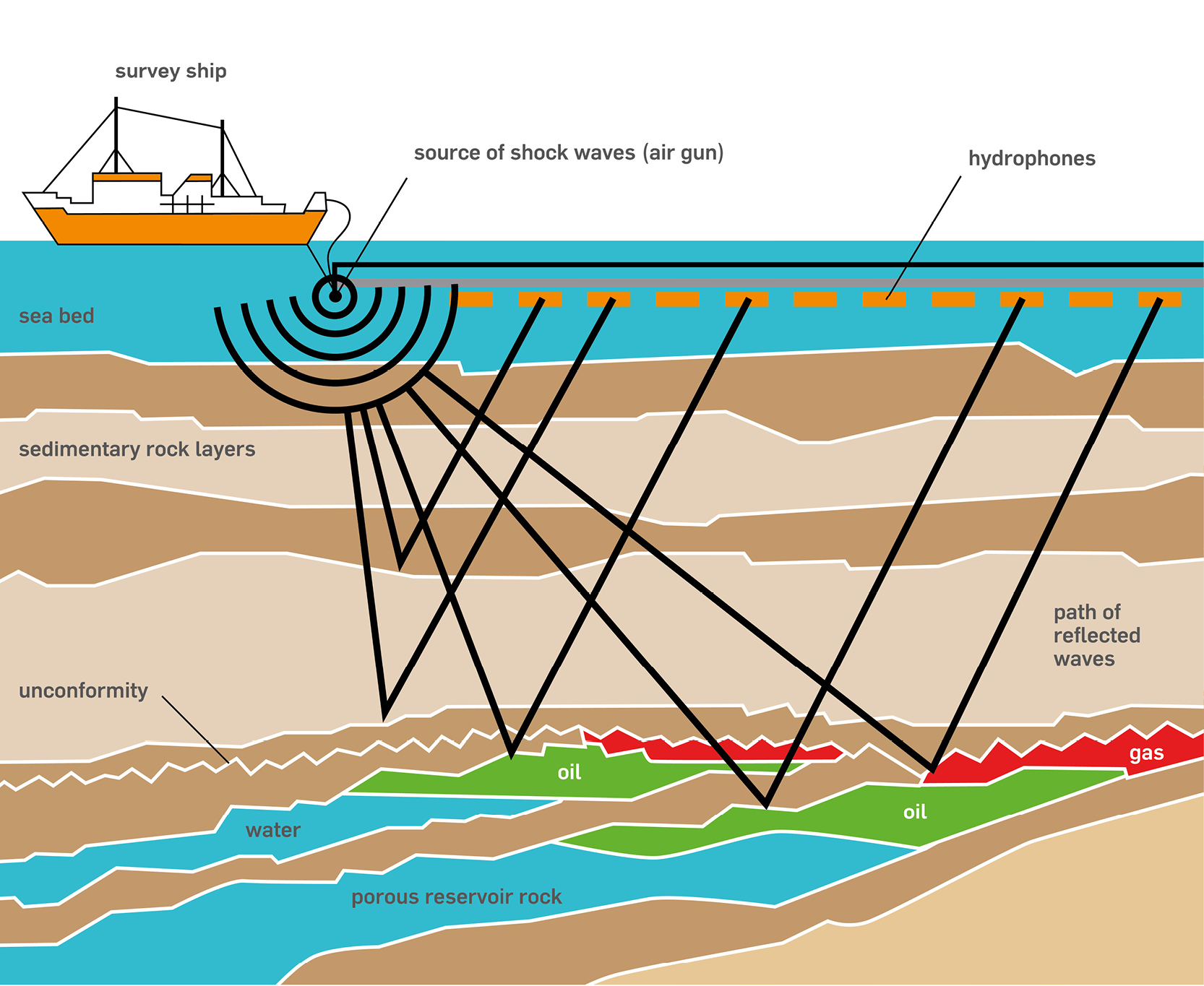}
\caption{Graphical demonstration of a seismic experiment that produces
  the data used as input in a seismic imaging workflow (Source: Open
  University \cite{openu})}
  \label{fig:seismic_survey}
\end{figure}

Seismic imaging techniques exploit the principle that a traveling
wave carries information about the physical properties of the
medium it travels through. While different techniques focus on
different kinds of information and objectives, we focus here on
reverse-time migration (RTM, \cite{Tarantola, Baysal1514}), an imaging
method that relies on a good estimate of the velocity model to obtain
an image of the reflectors in the subsurface. The algorithm relies on
a data-fitting procedure where synthetic data $\mathbf{d}_{syn}$ is
computed with the current estimate of the physical model via a
wave-equation solve and compared to the field measured data
$\mathbf{d}_{obs}$. An example of a field data recording is
illustrated in Figure \ref{fig:seismic_survey}. This problem is a
least-square minimization. We introduce here the formulation of the
problem solved and justify the implementation of optimal checkpointing. 
We previously introduced Devito \cite{kukreja2016devito}, a finite-difference domain specific language (DSL) for
time-dependent PDE solvers. Devito provides symbolic abstractions to
define the forward and adjoint wavefields. We will not go through its
implementation here but concentrate on the computation of the image of 
the subsurface.

Seismic imaging, in our case reverse-time migration (RTM), provides an image of the
subsurface reflectors from field recorded data and a cinematically
correct smooth background velocity model. In practice, the recording
is repeated with different source/receivers pair (called experiments)
over the same physical region. An estimate of the physical parameters $\mathbf{m}$ is
obtained from the recorded data with different methods such as
full-waveform inversion (FWI, iterative RTM
for low frequencies). Once $\mathbf{m}$ is estimated, RTM
provides an image of the subsurface to be interpreted. As just stated,
RTM is a single gradient of the FWI objective that can be written as
\cite{haber10TRemp, virieux2009overview, plessix2006review}:
\begin{equation}
\label{eq:fwibasic}
\mathop{\hbox{minimize}}_{\mathbf{m}} \Phi_s(\mathbf{m})=\frac{1}{2}\left\lVert\mathbf{d}_{syn} - \mathbf{d}_{obs}\right\rVert_2^2
\end{equation}

The square slowness model $\mathbf{m}$ is a physical property of the
medium through which the wave is propagating. 
The gradient of the objective function $\Phi_s(\mathbf{m})$ with
respect to the square slowness $\mathbf{m}$ is given by: 
\begin{equation}
\label{eq:Grad}
 \nabla\Phi_s(\mathbf{m})=\sum_{t =1}^{n_t}\mathbf{u}[\mathbf{t}] \mathbf{v}_{tt}[\mathbf{t}] =\mathbf{J}^T\delta\mathbf{d}
\end{equation}
where $\delta\mathbf{d} = \mathbf{d}_{syn}- \mathbf{d}_{obs}$ is the data
residual, $\mathbf{J}$ is the Jacobian of the forward operator,
$\mathbf{u}$ is the forward wavefield and $\mathbf{v}_{tt}$ is the second-order time derivative of the adjoint wavefield. 

It can be seen that the evaluation of the gradient first requires the
simulation of the forward and adjoint wavefields. This is achieved by
modeling the wave equation using a discretization, usually finite
difference. Various forms of the wave equation exist, e.g. acoustic
isotropic, anisotropic - VTI/TTI and elastic. Each of these models the
physics to different levels, with corresponding levels of
complexity. Here, we focus on the acoustic equation although the
analysis applies to all the forms of the equation mentioned. 

In the discrete form, the acoustic wave equation from
\citep{kukreja2016devito} can be written as the following linear
system:
\begin{equation}
\label{eq:wavedisc}
\mathbf{A}(\mathbf{m}) \mathbf{u} = \mathbf{P}_s^T \mathbf{q}
\end{equation}
where $\mathbf{A}$ is the discretized wave-equation and $\mathbf{P}_s$
is the source-restriction operator. The wavefield $\mathbf{u}$ is then given by:
\begin{equation}
\label{eq:wavefield}
\mathbf{u} = \mathbf{A}^{-1}(\mathbf{m}) \mathbf{P}_s^T \mathbf{q}
\end{equation}
Although equation \ref{eq:wavefield} can provide the value of $\mathbf{u}$ for
the entire domain at every time step, explicitly formulating the
entire matrix $\mathbf{u}$ is prohibitively expensive in terms of computer
memory required and is avoided wherever possible. When doing
forward-only simulations, of interest is the value of $\mathbf{u}$ at
certain predetermined locations in the simulated domain that we call
receivers. We record the progression of $\mathbf{u}$ at these
locations through time. This is represented mathematically by applying
the restriction operator $\mathbf{P}_{r}$ at the required receiver
locations. The result of applying $\mathbf{P}_{r}$ to $\mathbf{u}$,
which we call the \emph{simulated data}, is given by:
\begin{equation}
\label{eq:restricted}
\mathbf{d}_{syn} = \mathbf{P}_{r} \mathbf{A}^{-1}(\mathbf{m}) \mathbf{P}_s^T \mathbf{q}
\end{equation}

We can now rewrite the objective function from
equation~\ref{eq:fwibasic} as:
\begin{equation}\label{eq:AS}
	\mathop{\hbox{minimize}}_{\mathbf{m}} \Phi_s(\mathbf{m})=\frac{1}{2}\left\lVert\mathbf{P}_r
	\mathbf{A}^{-1}(\mathbf{m})\mathbf{P}_s^T\mathbf{q} - \mathbf{d}_{obs}\right\rVert_2^2
\end{equation}

In equation~\ref{eq:Grad} we can define the Jacobian of the forward
operator as:
\begin{equation}
J=\frac{d\mathbf{P}_{r} \mathbf{A}^{-1}(\mathbf{m}) \mathbf{P}_s^T}{d\mathbf{m}}
\end{equation}
While the $\mathbf{u}$ term in this equation can be calculated from
equation~\ref{eq:wavefield}, the $\mathbf{v}$ term can be calculated
from its adjoint equation given as:
\begin{equation}
\label{eq:adjoint}
 \mathbf{A}^T(\mathbf{m}) \mathbf{v} = \mathbf{P}_r^T \delta\mathbf{d} .
\end{equation}

\subsection{Implementation}
As the forward wavefield is obtained as a time-marching procedure
forward in time, the adjoint wavefield is then obtained similarly with a
backward in time time-marching procedure. The procedure to derive an
image with RTM can be summarized as:
\begin{enumerate}
	\item Compute the synthetic data $\mathbf{d}_{syn}$ with a forward solve with equation \eqref{eq:restricted}.
	\item Compute the adjoint wavefield from the data residual with equation \eqref{eq:adjoint}.
	\item Compute the gradient as the correlation of the forward and adjoint wavefield with equation \eqref{eq:Grad}.
\end{enumerate}

For the first step, we know from \citep{lange2017optimised} that 
equation~\ref{eq:wavefield} can be modeled in Devito using the
\lstinline|Operator| defined in figure~\ref{fig:fwd_code}
\lstset{language=Python}

\begin{figure}
\begin{lstlisting}
def forward(model, m, eta, src, rec, order=2): 
    # Create the wavefield function
    u = TimeData(name='u', shape=model.shape, time_order=2, space_order=order)
    
    # Derive stencil from symbolic equation
    eqn = m * u.dt2 - u.laplace + eta * u.dt
    stencil = solve(eqn, u.forward)[0]
    update_u = [Eq(u.forward, stencil)]
    
    # Add source injection and receiver interpolation
    src_term = src.inject(field=u, expr=src * dt**2 / m)
    rec_term = rec.interpolate(expr=u)
    
    # Create operator with source and receiver terms
    return Operator(update_u + src_term + rec_term, subs={s: dt, h: model.spacing})
\end{lstlisting}
\caption{Devito code required for a forward operator \label{fig:fwd_code}}
\end{figure}

The \lstinline|Operator| thus created can be used to model a forward
where the \lstinline|src_term| contains the source to be injected into the field
and \lstinline|rec_term| will be extracting the receiver information
from the simulation. 

Clearly, the third step requires the intermediate data from both the
previous steps. Storing both the forward and adjoint wavefields in
memory would be naive since two complete wavefields need to be
stored. The first obvious optimization is to merge steps 2 and 3 as a
single pass where step 3, the gradient calculation, can use the output
from step 2 (the adjoint wavefield) as it is calculated, hence saving
the need for storing the adjoint wavefield in memory. 
The \lstinline|Operator| for the combined steps 2 and 3 can be created
in Devito using the code given in figure~\ref{fig:grad_code}.
\begin{figure}
\begin{lstlisting}
def gradient(model, m, eta, src, rec, order=2):
    # Create the adjoint wavefield function
    v = TimeData(name='v', shape=model.shape, time_order=2, space_order=order)
    
    gradient_update = Eq(grad, grad - u.dt2 * v)
    
    # The adjoint equation
    eqn = m * v.dt2 - v.laplace - eta * v.dt
    stencil = solve(eqn, v.backward)[0]
    eqn = Eq(v.backward, stencil)

    # Add expression for receiver injection
    ti = v.indices[0]
    receivers = rec.inject(field=v, expr=rec * dt * dt / m)

    return Operator([eqn] + [gradient_update] + receivers, subs={s:dt, h: model.get_spacing()}, time_axis=Backward)
\end{lstlisting}
\caption{Devito code required for an operator that calculates the
  adjoint and gradient in a single pass \label{fig:grad_code}}
\end{figure}

This still leaves the requirement of having the result of
step 1 available. The most efficient, from a computational point of
view, would be to store the full history of the forward wavefield
during the first step. However, for realistically sized models, it
would require TeraBytes of direct access memory. One solution would be
to store the field on disk but would lead to slow access memory usage
making it inefficient. This memory limit leads to checkpointing,
storing only a subset of the time history, then recomputing it during
the adjoint propagation. \revolve provides an optimal schedule for
checkpointing to store for a given model size, number of time steps
and available memory. The next section discusses how checkpointing is
implemented. 

\section{Revolve}
\label{sec:revolve}
\medskip

\noindent
As we have seen, the usage of adjoint methods allows the computation of gradient information within
a time that is only a very small multiple of the time needed to evaluate the underlying 
function itself. However, for nonlinear processes like the one we saw
in the previous section, the memory requirement to compute the adjoint information
is in principle proportional to the operation count of the underlying
function, see, e.g., \cite[Sec.~4.6]{GrWa08}. In Chap.~12 of the same
book, several checkpointing alternatives to reduce this high memory
complexity are discussed. Checkpointing strategies use a small number
of memory units (checkpoints) to store the system state at distinct
times. Subsequently, the recomputation of information that is needed
for the adjoint computation but not available is performed using these
checkpoints in an appropriate way. Several checkpointing techniques
have been developed, all of which seek an acceptable compromise between
memory requirement and runtime increase. Here, the obvious question is
where to place these checkpoints during the forward integration to
minimize the overall amount of required recomputations.

To develop corresponding optimal checkpointing strategies, one has to
take into account the specific setting of the application. A fixed
number of time steps to perform and a constant computational cost of
all time steps to calculate is the simplest situation. It was shown in
\cite{griewank2000} that, for this case, a checkpointing scheme based on
binomial coefficients yields, for a given number of checkpoints, the
minimal number of time steps to be recomputed. An obvious extension of
this approach would be to include flexibility with respect to the
computational cost of the time steps. For example, if one uses an
implicit time stepping method based on the solution of a
nonlinear system, the number of iterations needed to solve the
nonlinear system may vary from time step to time step, yielding
non-uniform time step costs. In this situation, it is no longer
possible to derive an optimal checkpointing strategy
beforehand. Some heuristics were developed to tackle this
situation \cite{StHi10}. However, extensive testing showed that, even
in the case of nonuniform step costs, binomial checkpointing is quite
competitive. Another very important extension is the coverage of
adaptive time stepping. In this case, the number of time steps to be
performed is not known beforehand. Therefore, so-called online
checkpointing strategies were developed, see, e.g.,
\cite{StWa10,Wang:2009aa}. Finally, one has to take into account where
the checkpoints are stored. Checkpoints stored in memory can be lost
on failure. For the sake of resilience or because future
supercomputers may be memory constrained, checkpoints may have to
necessarily be stored to disk. Therefore, the access time to read or
write a checkpoint is not negligible in contrast to the assumption
frequently made for the development of checkpointing approaches. There
are a few contributions to extend the available checkpointing
techniques to a hierarchical checkpointing, see, e.g., \cite{Auetal16,Schetal15,StWa09}.

The software {\sf revolve} implements binomial checkpointing, 
online checkpointing as described in \cite{StWa10}, and 
hierarchical, also called multi-stage, checkpointing derived in
\cite{StWa09}. For this purpose, it provides a data structure {\sf r}
to steer the checkpointing process and the storage of all information
required for the several checkpointing strategies.

To illustrate the principle structure of an adjoint computation using
checkpointing, Fig.~\ref{fig:revolve} illustrates the kernel of {\sf
  revolve} used for the binomial checkpointing. The two remaining
checkpointing strategies are implemented in a similar fashion only
taking the additional extensions into account. The forward integration
as well as the corresponding adjoint computation is performed within a
{\sf do-while}-loop of the structure in Fig.~\ref{fig:revolve}, where
{\sf steps} and {\sf snaps} denote the number $n_t$ of time steps of the
forward simulation and the number $c$ of checkpoints available for the
adjoint computation, respectively.

\lstset{language=C++}
\begin{figure}
\begin{lstlisting}
r=new Revolve(steps,snaps)
do
    whatodo = r->revolve() 
    switch(whatodo)
        case advance: for r->oldcapo < i <= r->capo 
                                   forward(x,u)
        case firsturn:  eval_J(x,u)
                          init(bu,bx)
                          adjoint(bx,bu,x,u)
        case youturn: adjoint(bx,bu,x,u)
        case takeshot: store(x,xstore, r->check)
        case restore: restore(x,xstore, r->check)
while(whatodo <> terminate)
\end{lstlisting}
\caption{ {\sf revolve} algorithm with calls to the application interface \label{fig:revolve}}
\end{figure}
\lstset{language=Python}
Hence, the routine {\sf revolve} determines the next action to be
performed which must by supported by the application being
differentiated. These actions are
\begin{itemize}
\item \textbf{advance}: Here, the user is supposed to perform  a part of the
  forward integration based on the routine {\sf forward(x,u)}, where
  {\sf x} represents the state of the system and {\sf u} the
  control. The variable {\sf r--$>$oldcapo} contains the current
  number of the state of the forward integration. That is, before
  starting the for-loop $x$ holds the state at time $t_{\mbox{\sf
      r--$>$oldcapo}}$. The variable {\sf r--$>$oldcapo} determines
  the targeted number of the state  of the forward
  integration. Therefore, {\sf r--$>$capo} $-$ {\sf r--$>$oldcapo}
  time steps have to be perform to propagate the state $x$ from the
  time $t_{\mbox{\sf r--$>$oldcapo}}$ to the time $t_{\mbox{\sf
      r--$>$capo}}$ 
\item \textbf{firstrun}: This action signals the start of the adjoint
  computation. Therefore, first the target function is
  evaluated. Then, the user has the possibility to initialize the
  adjoint variable {\sf bu} and {\sf bx}. Subsequently, the first
  adjoint step is performed.  
\item \textbf{youturn}: The next adjoint step has to be performed.
\item \textbf{takeshot}: Here, the  user is supposed to store the current state
  $x$ in the checkpoint with the number {\sf r--$>$check}. The array
  of checkpoints is here denoted by {\sf xstore} but the specific
  organisation of the checkpoints is completely up to the user. During
  the adjoint computation {\sf r--$>$check} selects the checkpoint
  number appropriately such that all states needed for the adjoint
  computation are available. Once the adjoint computation has started,
  states that were stored in the checkpoints are also overwritten to
  reuse memory. 
\item \textbf{restore}: The content of the checkpoint with the number {\sf
    r--$>$check} has to be copied into the state $x$ to recompute the
  forward integration starting from this state. 
\end{itemize}
It is important to note that this checkpointing approach is completely
independent from the method that is actually used to provide the
adjoint information. As can be seen, once an adjoint computation
is available the implementation can incorporate 
binomial checkpointing to reduce the memory requirement.

We also have to stress that {\sf revolve} provides a so-called serial
checkpointing which means that only one forward time step or one
adjoint step is performed at each stage of the adjoint
computation. Nevertheless, the computation of the forward time step
and/or the adjoint step may be performed heavily in parallel, i.e.,
may be evaluated on a large scale computer system. This is in contrast
to so-called parallel checkpointing techniques where several forward
time steps might be performed in parallel even in conjunction with one
adjoint step. Corresponding optimal parallel checkpointing schedules
were developed in \cite{Wa04}. However, so far no implementation to
steer such a parallel checkpointing process is available.

The {\sf revolve} software also includes an \textbf{adjust} procedure that
computes, for a given number of time steps, the number of checkpoints such
that the increase in spatial complexity equals approximately the
increase in temporal complexity. Using the computed number as the
number of checkpoints minimises cost when assuming that the user pays
computational resources per node and per time, e.g. the cost is
proportional to the available memory and the runtime of the
computation.

\section{Abstractions for Checkpointing}
\label{sec:api}
\lstset{language=Python}
In this section, we will discuss the package \pyrevolve, which has been
developed during the course of this work to encapsulate Revolve
checkpointing in a user-friendly, high level python library. This
library is available online, along with its source \footnote{https://github.com/opesci/pyrevolve}. We will
first provide details of its implementation in
Section~\ref{sec:apipy}. Afterwards, we will discuss the interplay of
\pyrevolve with the C++ checkpointing implementation that was
previously discussed in Section~\ref{sec:revolve}. Finally, we will
discuss the usage of \pyrevolve in an application in
Section~\ref{sec:apidev} to RTM, as described in
section \ref{sec:inversion_devito}, implemented in the Devito domain specific language.

Although section \ref{sec:apidev} discusses the special case of
devito, the interface of the \pyrevolve library was designed to allow
an easy integration into other python codes as well.

\subsection{API, pyRevolve side}
\label{sec:apipy}
The \pyrevolve interface was designed as part of providing
checkpointing to users of Devito with an accessible API. The design
had the following goals:
\begin{itemize}
\item Making checkpointing available to users of Devito without
  forcing them to get involved in implementation details like loops, callbacks, data storage
  mechanisms. The user should choose whether to use checkpointing in
  one place, but not be forced to do anything beyond this.
\item All knowledge of checkpointing, different strategies
  (online/offline checkpointing, multistage) shall be contained within
  one module of the python framework, while still benefiting all operations in the code.
\item The checkpointing itself should be contained in a separate
  library that allows others to use it easily, even if they are not
  interested in using Devito. This matured into \pyrevolve. 
\item Since the data movement requires intricate knowledge of the data
  structures used and their organization in memory, this is handled by
  the application code.
\end{itemize}

To achieve these goals, \pyrevolve was designed for the following
overall workflow, which will be explained in more detail in the
following sections. The term \emph{application} here refers to
the application using \pyrevolve as a library (in this case Devito).
To begin with, the application creates objects with an \emph{apply} method to perform the
  actual forward and reverse computations, which are both instances of
  a concrete implementation of the abstract base class\emph{Operator}.
The application also creates an instance of a concrete implementation
of the abstract base class \emph{Checkpoint} that can deep-copy all time-dependent
  working data that the operators require into a specified memory
  location. Next, the application instantiates \pyrevolve's \emph{Revolver} object and
  passes the forward and reverse operators, and the checkpoint object.
When required, the application starts the Revolver's forward sweep, which will
  complete the forward computation and store checkpoints as
  necessary. After the forward sweep completes, the application can
  finalize any computation that is based on the forward data, such as
  evaluation of objective functions, or store the final result as
  necessary. This may be accompanied/followed by the initialization of the adjoint data structures.
After this, the application calls the Revolver's reverse sweep. This will compute
  the adjoint, possibly by performing partial forward sweeps and
  loading checkpoint data.

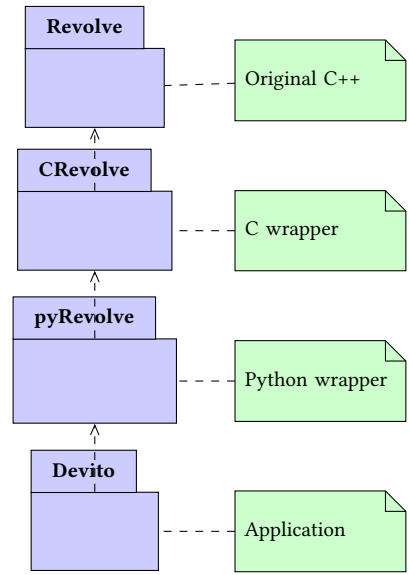
\begin{figure}
\begin{tikzpicture}
\begin{umlemptypackage}[x=0, y=-0.1]{Revolve}\end{umlemptypackage}
\umlnote[x=3, y=0, width=2cm]{Revolve}{Original C++}

\begin{umlemptypackage}[x=0, y=-2]{CRevolve}\end{umlemptypackage}
\umlnote[x=3,y=-2, width=2cm]{CRevolve}{C wrapper}

\begin{umlemptypackage}[x=0, y=-4]{pyRevolve}\end{umlemptypackage}
\umlnote[x=3,y=-4, width=2cm]{pyRevolve}{Python wrapper}

\begin{umlemptypackage}[x=0, y=-6]{Devito}\end{umlemptypackage}
\umlnote[x=3,y=-6, width=2cm]{Devito}{Application}

\umldep[anchors=90 and 270]{CRevolve}{Revolve}
\umldep[anchors=90 and 270]{pyRevolve}{CRevolve}
\umldep[anchors=90 and 270]{Devito}{pyRevolve}

\end{tikzpicture}
\caption{Packages overview. Devito and an example application that uses checkpointing are the subject of Section~\ref{sec:inversion_devito}. Revolve has been described in Section~\ref{sec:revolve}. The packages \pyrevolve and \crevolve and how they are used to create a high-level abstraction of checkpointing are explained in Section~\ref{sec:api}.}
\label{fig:UMLPackages}
\end{figure}

The \pyrevolve package contains crevolve, which is a thin C wrapper around a
previously published C++
implementation\footnote{http://www2.math.uni-paderborn.de/index.php?id=12067\&L=1}. The
C++ files in this package are slightly modified for compatibility with Python, but the
original is available from the link in the footnote. The crevolve wrapper around the
C++ library is taken from
libadjoint\footnote{https://bitbucket.org/dolfin-adjoint/libadjoint}.

One key design aspect is that \pyrevolve is not responsible for performing the data copies, and
therefore does not need to know about the properties or structure of the data that needs to be
stored. For this purpose, \pyrevolve provides the \texttt{Checkpoint} abstract base class that has a
\texttt{size} attribute, and a \texttt{load(ptr)} and \texttt{save(ptr)} method. The user must
provide a concrete implementation of such an object. The \texttt{size} attribute must contain the
size of a single checkpoint in memory, this information is used by \pyrevolve to allocate the correct
amount of memory.  The \texttt{save} method must deep-copy all working data to the memory region
starting at the provided pointer, and the \texttt{load} method must restore the working data from
the memory region starting at the pointer \texttt{ptr}, either by performing a deep-copy, or by
pointing the computation to the existing data inside the checkpoint storage.

The \pyrevolve library provides to the user the class \texttt{Revolver} that must be instantiated with the following arguments:

\begin{itemize}
\item \textbf{Checkpoint object:} This has to be an implementation of the abstract base class
\texttt{Checkpoint}.
\item \textbf{Forward operator:} An object that provides a function \texttt{apply()} as specified in
Section~\ref{sec:apidev} that performs the forward computation.
\item \textbf{Reverse operator:} Similarly, an object that provides a function \texttt{apply()} that
performs the reverse computation.
\item \textbf{Number of checkpoints:} This is optional, and specifies the number of checkpoints that
can be stored in memory. If it is not given, a default value is computed using the \texttt{adjust}
method explained in Section~\ref{sec:revolve}.
\item \textbf{Number of time steps:} This is also optional. If it is not given, an online
checkpointing algorithm is used.
\end{itemize}

Based on either the given or computed number of checkpoints, the constructor instantiates a storage
object that allocates the necessary amount of memory (\textit{number of checkpoints} $\times$
$checkpoint size$), and makes that memory accessible to the \texttt{Revolver}.

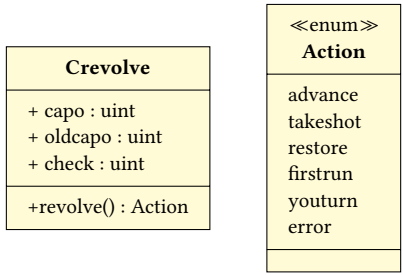
\begin{figure}
\begin{tikzpicture}
\umlclass[x=1.5, y=0, type=enum]{Action}{
  advance\\
  takeshot\\
  restore\\
  firstrun\\
  youturn\\
  error
}{}

\umlclass[x=-1.5, y=0]{Crevolve}{
+ capo : uint\\
+ oldcapo : uint\\
+ check : uint
}{
+revolve() : Action
}
\end{tikzpicture}
\caption{Crevolve classes.}
\label{fig:UMLCrevolve}
\end{figure}

\begin{figure}
\begin{tikzpicture}
\umlclass[x=-1.5, y=0, type=abstract]{Checkpoint}{
\umlvirt{\# size : int}
}{
\umlvirt{\# save(ptr : pointer) : void}\\
\umlvirt{\# load(ptr: pointer) : void}
}

\umlclass[x=1, y=-2.7]{Storage}{
- data : numpy\_array
}{
{+ init(n\_ckp : int, size\_ckp : int)}\\
{+ get\_item(i : uint) : pointer}
}

\umlclass[x=0, y=-6.3]{Revolver}{
- op\_forward : Operator\\
- op\_reverse : Operator\\
- store : Storage\\
- cr : CRevolve :: Checkpointer\\
- ckp : Checkpoint
}{
{+ init(ckp, op\_forward, op\_reverse, n\_ckp, n\_time)}\\
{+ apply\_forward() : void}\\
{+ apply\_reverse() : void} 
}

\umlclass[x=0, y=-10, type=abstract]{Operator}{
}{
{\umlvirt{\# apply() : void}}
}

\umluniassoc[mult=1, anchors=47 and 308,
loopsize=2cm]{Revolver}{Storage}

\umlcompo[mult=1, anchors=140 and 230,
loopsize=2cm]{Revolver}{Checkpoint}

\end{tikzpicture}
\caption{\pyrevolve classes. The abstract classes \texttt{Checkpoint}
  and \texttt{Operator} are implemented by the client application}
\label{fig:UMLpyrevolve}
\end{figure}
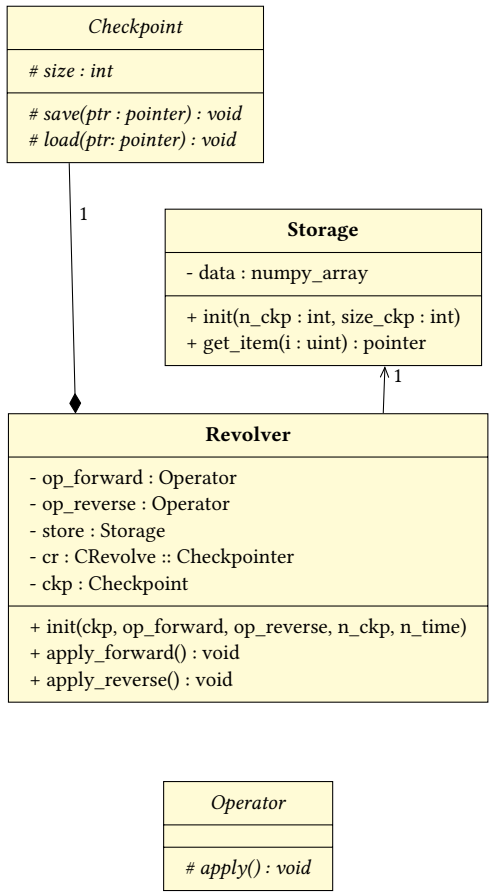

\subsection{API, Application side}
\label{sec:apidev}

To introduce checkpointing using the \pyrevolve library, an application must implement a particular
interface. We use devito here as an example application, however,
everything discussed here is fairly general and any application may
implement checkpointing using \pyrevolve using the approach discussed
in this section. 

To begin with, a concrete implementation of \pyrevolve's abstract base
class \emph{Checkpoint}, called \emph{DevitoCheckpoint} was
created. This class has three methods:
\begin{itemize}
\item \textbf{save(ptr)}: Save the contents of the working memory into
  the location \lstinline|ptr|. 
\item \textbf{restore(ptr)}: Restore a previously stored checkpoint
  from location \lstinline|ptr| into working memory. 
\item \textbf{size}: Report the amount of memory required by a single
  checkpoint. This is used to decide the total amount of memory to be
  allocated and to calculate offsets. 
\end{itemize}

Along with this, we set up two \lstinline|Operator|, one a \lstinline|ForwardOperator| to carry out the
forward computation and a \lstinline|GradientOperator| that computes
the image, as explained in
section~\ref{sec:inversion_devito}. These can be used to initialize a
\lstinline|Revolver| object as shown in
figure~\ref{fig:devito_pyrevolve}. 

\begin{figure}
\begin{lstlisting}
# Example time-varying field that needs checkpointing
u = TimeData(...)
# Some expression that generates the values for u
fw = Operator(...)
# Some expression that uses the values of u
rev = Operator(...)
cp = DevitoCheckpoint([u])
revolver = Revolver(cp, fw, rev, nt)
# Forward sweep that will pause to take checkpoints
revolver.apply_forward() 
# Could perform some additional steps here
# Reverse sweep that uses the checkpoints
revolver.apply_reverse() 
\end{lstlisting} 
\caption{Devito code to utilize checkpointing based on \pyrevolve \label{fig:devito_pyrevolve}}
\end{figure}
On initialization of the \lstinline|Revolver|, object, the
\lstinline|DevitoCheckpoint| object is queried for the size of one
checkpoint and \pyrevolve allocates
\lstinline|n_checkpoints*checkpoint.size| bytes of memory for the
storage of checkpoints. 
Calling \lstinline|revolver.apply_forward()| carries out a forward run
but broken down into chunks as specified by the checkpointing schedule
provided by \emph{Revolve}, each chunk being executed by calling
\lstinline|fwd.apply()| with arguments \lstinline|t_start| and
\lstinline|t_end| corresponding to the timesteps to run the simulation
for. Between these chunks, \lstinline|cp.save()| is automatically
called to save the state to a checkpoint. 

\begin{sloppypar}
On calling \lstinline|revolver.apply_reverse()|, the
\lstinline|Revolver| calls \lstinline|rev.apply()| with the relevant
\lstinline|t_start|/\lstinline|t_end| arguments for the sections where
the result from the forward pass is available in a checkpoint. This
will be loaded by a call to \lstinline|cp.load|. For
others, it will automatically call \lstinline|fwd.apply()| to
recompute and store in memory the results from a part of the forward
operator so the reverse operator can be applied for that part. 
\end{sloppypar}

\begin{figure}
\begin{tikzpicture}
\umlclass[x=0, y=0]{DevitoCheckpoint}{
{+ size : int}
}{
{+ save(ptr : pointer) : void}\\
{+ load(ptr: pointer) : void}
}

\umlclass[x=-2, y=-4.3]{ForwardOperator}{
- data
}{
{+ apply() : void}
}

\umlclass[x=2, y=-4.3]{ReverseOperator}{
- data
}{
{+ apply() : void}
}

\end{tikzpicture}
\caption{Devito operators, and the implementation of a \texttt{Checkpoint} class.}
\label{fig:UMLDevito}
\end{figure}

Following this work, users of Devito can easily add optimal checkpointing to their adjoint
computations by following the steps described above.

\section{Experiment}
\label{sec:experiment}
\begin{figure}
\label{fig:velocitymodel}
\includegraphics[width=0.8\linewidth]{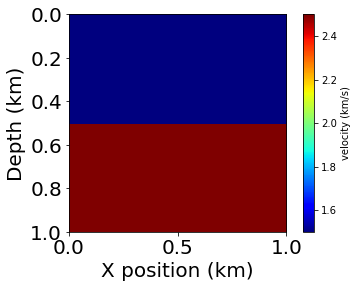} 
\caption{Velocity model used for gradient test}
\end{figure}

There are two possible ways of testing the numerical accuracy of an
implementation - solving a problem whose solution has certain known
mathematical properties and verifying these properties numerically,
and comparing the results to a reference solution. Here we do both -
we use the gradient test as described in \citep{kukreja2016devito} and
also verify that the numerical results match those from a reference
implementation. The test uses the Taylor property of the gradient to
test whether the calculated gradient follows the expected convergence
for small perturbations. The test can be written mathematically as:
\begin{align}
 \epsilon_0 = &\Phi_s(\mathbf{m_0} + h \mathbf{dm}) - \Phi_s(\mathbf{m_t})\nonumber\\
 \epsilon_1 = &\Phi_s(\mathbf{m_0} + h \mathbf{dm}) - \Phi_s(\mathbf{m_0}) - h \langle\nabla\Phi_s(\mathbf{m_0}),\mathbf{dm}\rangle.
\label{GrFWI:test}
 \end{align}

where $\Phi_s$ is defined in equation~\ref{eq:fwibasic}. This test is
carried out for a certain $\mathbf{m}$, which we call here
$\mathbf{m_0}$. This is the smoothed version of a two layer model
$\mathbf{m_t}$ i.e. the true model has two horizontal sections, each
with a different value of squared
slowness. Figure~\ref{fig:velocitymodel} shows the true velocity model
$\mathbf{m_t}$. The measured data required by the objective function
is modeled on the true two-layer model and $\mathbf{dm} = \mathbf{m_0}
- \mathbf{m_t}$. The constant $h$ then varies between $10^{-1}$ and
$10^{-4}$ to verify that $\epsilon_0 = O(h)$ is a first order error and
$\epsilon_1 = O(h^2)$ is a second order error. The code used for this
test can be found in the repository for devito
\footnote{https://github.com/opesci/devito}. The tests were carried
out on a Intel(R) Xeon(R) CPU E5-2640 v3 @ 2.60GHz (Haswell) with
128GB RAM.

We used a grid with $230 \times 230 \times 230$ points. With the
simulation running for 1615 timesteps, this required about 80 GB to
store the full forward wavefield in memory. The first run
was made with the regular gradient example that stores the entire
forward wavefield in memory. This was then repeated with
checkpointing, with varying number of checkpoints. It was verified
that the results from all versions matched each other exactly and also
passed the gradient test mentioned previously. The peak memory
usage was tracked for each such run, as well as the total time to
solution. The memory consumption was measured using the
\emph{memory\_profiler} python module and the time to solution by using
the \emph{time} python command before and after the function to be
profiled. To eliminate variation in the results, the timings are the
minimum value from three runs. 

\begin{figure}
\includegraphics[width=0.8\linewidth]{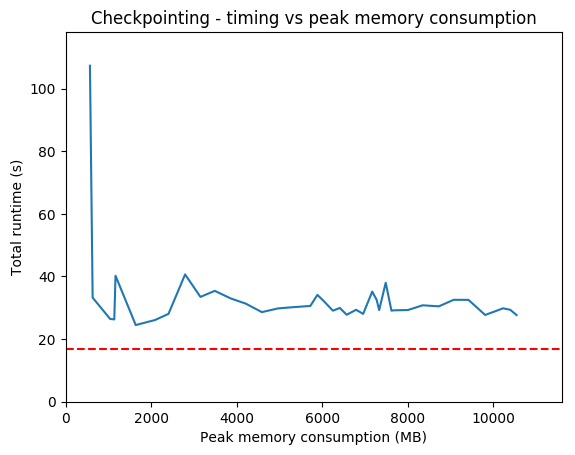} 
\caption{Timings for gradient test for different amounts of peak
  memory consumption}\label{fig:timings}
\end{figure}

As can be seen in figure~\ref{fig:timings}, the reduction in runtime
as more memory is available is not in line with the theoretical
predictions from \citet{griewank2000}. This is expected because the
theoretical numbers do not take into account the cost of deep copies
(implemented here using \emph{numpy}) as well as the cost of
repeatedly calling a C function from python vs doing the repetition
inside the C function. Since the adjoint computation (as well as the
associated forward computation) is carried out one time-step at a
time, this significantly reduces the amount of work available inside a
single \lstinline|Operator| call. This might cause inefficiencies in
load-balancing across cores using OpenMP. This effect is seen most
clearly when comparing the reference implementation that stores the
forward field in a contiguous block of memory with the checkpointed
implementation that stores a checkpoint at every time step. In this
case, although the memory consumption of the two implementations is
the same, the checkpointed implementation runs slower because of the
overheads previously mentioned. 
\section{Conclusions}
Through this work, we have shown that with high-level abstractions it
is possible to greatly simplify the complexity of client code. We have
also verified the correctness of our implementation using mathematical
tests. This already enables the users of Devito to utilize Revolve
based checkpointing in their applications to solve much bigger
problems than previously possible. However, through the experiment in
section~\ref{sec:experiment}, we have seen that the overhead
introduced by checkpointing is non-trivial. For this reason, there is
much more work to be done to implement more features that widen the
applicability of \pyrevolve.

\section{Future Work}
This work carried out so far was a proof of concept of integration
with Revolve using high level abstractions and, as such, is still a
work-in-progress in terms of use for practical applications. The most
important limitation in the current implementation is that it
implements ``serial checkpointing'', i.e. during the reverse
computation, only one timestep can be advanced at a time and this
severely limits the parallelizability of this code. The high-level
interface would need to be extended to be able to manage
parallelization strategies. Another important feature that might be
required in \pyrevolve before it is adopted in the community is
multi-stage checkpointing. Here, some checkpoints may be transparently
swapped to disk, further increasing the amount of memory available to
applications without any change in the application code. 
For problems implemented with adaptive time-stepping, the number of
time-steps is not known a-priori and that would require \pyrevolve to implement
online checkpointing, something that even Devito would require in
future versions. 

\begin{acks}
  The authors are very grateful to Fabio Luporini and Nicolas Barral. This work
  was carried out as part of the Intel Parallel Computing Centre at
  Imperial College, London. 
\end{acks}

\bibliographystyle{ACM-Reference-Format}
\bibliography{bib_checkpointing} 

\end{document}